\documentclass [11pt,a4paper] {article}

\usepackage[cp1252]{inputenc}
\usepackage[english]{babel}

\usepackage{amssymb}
\usepackage{amsmath}
\usepackage{amsfonts,amssymb}

\usepackage[dvips]{graphicx}

\begin{document}

\title{Justifying the Classical-Quantum Divide of the Copenhagen Interpretation}

\author{Arkady Bolotin\footnote{$Email: arkadyv@bgu.ac.il$} \\ \textit{Ben-Gurion University of the Negev, Beersheba (Israel)}}

\maketitle

\begin{abstract}\noindent Perhaps the most significant drawback, which the Copenhagen interpretation (still the most popular interpretation of quantum theory) suffers from, is the classical-quantum divide between the large classical systems that carry out measurements and the small quantum systems that they measure. So, an “ideal” alternative interpretation of quantum theory would either eliminate this divide or justify it in some reasonable way. The present paper demonstrates that it is possible to justify the classical-quantum dualism of the Copenhagen interpretation by way of the analysis of the time complexity of Schrödinger's equation.\\

\end{abstract}

\section{Introduction}

\noindent As stated by the standard Copenhagen position on quantum mechanics \cite{Faye}, microscopic systems under consideration are described by wave functions or state vectors, whose time evolution is governed by the Schrödinger equation, whereas observers can access those systems only through macroscopic measuring devices that (together with the observers themselves) are subjects to the laws of classical physics. In this way, the standard Copenhagen position postulates that the world is governed by different laws: quantum mechanics (explicitly, Schrödinger's equation) for the microscopic world, and classical physics (Newton's laws of motion) for the macroscopic, directly accessible, world.\\

\noindent The existence of two divided physical domains is considered by the many as the core weakness of the Copenhagen position \cite{Landsman}. Besides this, however, such an interpretation of quantum mechanics constitutes a coherent framework for the description of the physical world, which works quite well in most practical circumstances. So, at least among practicing physicists, the Copenhagen interpretation is still the most popular interpretation today \cite{Schlosshauer13}.\\

\noindent The burning question is – if one left the mathematical structure of the Copenhagen interpretation intact (since this structure is widely accepted), would be an alternative interpretation of quantum theory possible that either would eliminate the “absurd” classical-quantum dualism or would convincingly justify it?\\

\noindent A great deal of papers (for example \cite{Styer,Vaidman,Penrose}, just to name a few), which argue that the Schrödinger equation can flawlessly predict the future behavior of all physical systems microscopic and macroscopic alike (including observers represented by their own wave functions), proves that the first option – i.e., an interpretation without the classical-quantum dualism of the Copenhagen interpretation – is possible.\\

\noindent The aim of this paper is to demonstrate that the second option – i.e., the alternative quantum-mechanical interpretation that justifies the classical-quantum divide through the analysis of the time complexity of Schrödinger's equation – is possible as well.\\

\noindent The paper is structured as follows. First, we will consider the complexity of verification of exact solutions $\!\left.\left|\psi \!\right.\right\rangle$ to Schrödinger's equation $H\!\!\left.\left|\psi \!\right.\right\rangle\!=\!0$ and show that for a very wide class of non-relativistic many-body Hamiltonians $H$, the decision form of this equation (i.e., does this equation have a solution?) can be verified in the amount of time polynomial in the system's constituent particle number. Next, it will be demonstrated that the complexity of verification of the Schrödinger equation for coupled spin systems (characterized with the Hamiltonian, whose first part contains all the interactions described by the distances between constituent particles of the system, while the second part contains coupling to an external magnetic field as well as coupling between spins of the constituent particles) is polynomial as well. Then, it will be shown that if the Exponential Time Hypothesis held true, no generic algorithm capable of exactly solving the Schrödinger equation for an arbitrary physical Hamiltonian could be significantly faster than the brute-force procedure of generating and testing all possible candidate solutions of this equation. Finally, it will be shown that the key element of the Copenhagen interpretation, which postulates the necessity of classical concepts in order to describe quantum phenomena, including measurements, can be explained by NP-hardness of the problem of exactly solving the Schrödinger equation for a macroscopic system.\\

\section{Complexity of verification of Schrödinger's equation exact solutions}

\noindent Let us consider the family ${\Phi }_{\Psi }$ of generic algorithms capable of exactly solving the Schrödinger equation for an arbitrary physical Hamiltonian $H$. By “exactly solving” we mean that the algorithms $\in{\Phi }_{\Psi }$ can determine exactly (i.e., in exact, more or less closed form) all (or at least the first several lower) eigenvalues and corresponding eigenfunctions of a given Hamiltonian $H$, and by “generic” we mean that those algorithms can do it so for any and all possible physical systems with any possible numbers of constituent particles $N$. We know that the family ${\Phi }_{\Psi }$ is not empty and contains as a minimum one member: it is brute force, that is, the procedure of generating and testing all possible candidate solutions to the Schrödinger equation with a given Hamiltonian $H$. Let us assume that the family ${\Phi }_{\Psi }$ contains at least one more member, which we will call the algorithm $A\!\left({\Phi }_{\Psi }\!\right)$.\\

\noindent Suppose the vector $\!\left.\left|\psi \!\right.\right\rangle$ is the exact solution to the Schrödinger equation $H\!\!\left.\left|\psi \!\right.\right\rangle\!=\!0$ for the ground state of the particular Hamiltonian $H$ with zero eigenvalue $E\!=\!0$ found by the algorithm $A\!\left({\Phi }_{\Psi }\!\right)$. Let us show that the decision problem of this equation (i.e., does this equation have a solution?) can be verified quickly, i.e., in an amount of time polynomial in $N$. Clearly, the way to accomplish this is to substitute the solution $\!\left.\left|\psi \!\right.\right\rangle$ back into the Schrödinger equation with the given $H$ and estimate the runtime complexity of the operations needed to prove that $\!\left.\left|\psi \!\right.\right\rangle$ is indeed the solution.\\

\noindent Let $L$ be the minimal number of elementary operations sufficient to compute the effect of the Hamiltonian $H$ on the solution $\!\left.\left|\psi \!\right.\right\rangle$; we will call $L$ the complexity of verification. In the position basis $\{\!\left.\left|{\mathbf r}\!\right.\right\rangle\}$ the Hamiltonian $H$ is quadratic in the operators ${\partial}\!/{\partial {\mathbf r}_j}$, thus using the results of the papers \cite{Paterson,Baur} the complexity $L$ can be presented as follows:\\

\begin{equation} \label{1} 
   L\left(
        H\!\!\left.\left|\psi \!\right.\right\rangle 
   \right)
   =
   L\left(
         \frac{{\partial }^2\Psi }{\partial {{\mathbf r}}^2_1},\dots, 
         \frac{{\partial }^2\Psi }{\partial {{\mathbf r}}^2_j},\dots,
         \frac{{\partial }^2\Psi }{\partial {{\mathbf r}}^2_N}
   \right)
   \le
   O\!\left(N^2\right)
   \!\cdot\!
   {\mit cost}\!\left(\!\Psi\!\right)
\;\;\;\;  ,
\end{equation}
\smallskip

\noindent where only partial derivatives ${\partial}^2\Psi\!/{\partial {{\mathbf r}}^2_j}$ (computed via the chain rule using the known partial derivatives of the elementary functions that form the wave function $\Psi$) are considered contributed to the complexity of verification $L$ (as binary elementary operations whose both operands involve the function $\Psi$), while additions/subtractions and multiplications by arbitrary scalars are allowed for free, ${\mit cost}\!\left(\!\Psi\!\right)$ denotes the computational cost of the evaluation of the wave function $\Psi\!\left({\mathbf r},m\right)$ at the particular values of position vectors ${\mathbf r}=\left({{\mathbf r}}_1,\dots ,{{\mathbf r}}_j,\dots ,{{\mathbf r}}_N\right)$ and spin components along the $z$-axis $m=\left({m}_1,\dots ,{m}_j,\dots ,{m}_N\right)$.\\

\noindent In the case of a weak or zero magnetic field along the $z$-axis and non-coupled spins, the Schrödinger equation for the $N$-body system with a very general potential described by the distances $d_{jk}$ between the $j$ and the $k$ particles of the system (such as Coulomb, inverse square, harmonic-oscillator and dipole types or as the Yukawa potential, the Gauss potential, the negative exponential potential and so on) can be transformed using hyperspherical coordinates $\rho$ and $\Omega$, where $\rho$ is the hyperradius and $\Omega$ stands for the collective hyperspherical angles. Using this coordinate transformation, the exact solution $\Psi\!\left(\rho,\Omega\right)$ to the system's Schrödinger equation can be separated into the product of the hyperradial and hyperangular wave functions\\

\begin{equation} \label{2} 
   \Psi\!\left(\rho,\Omega\right)
   =
   \phi\!\left(\rho \right)\!\Upsilon\!\left(\Omega \right)
\;\;\;\;  ,
\end{equation}
\smallskip

\noindent and so the computational cost of the evaluation of the solution $\Psi\!\left(\rho,\Omega\right)$ can be estimated separately as\\

\begin{equation} \label{3} 
   {\mit cost}\!\left(\!\Psi\!\right)
   =
   {\mit cost}\!\left(\phi\right) + {\mit cost}\!\left(\Upsilon\right) + 1
\;\;\;\;  .
\end{equation}
\smallskip

\noindent There are different representations of the hyperspherical coordinates, but according to the paper \cite{ZhangR} calculations with them give the same results. Therefore, using a coordinate system in the $N$-dimensional configuration space analogous to the spherical coordinate system for 3-dimensional Euclidean space, we can estimate the numbers of elementary operations $L\!\left(\rho\right)$ and $L\!\left(\Omega_j\right)$ needed to calculate the hyperspherical coordinates $\rho$ and $\Omega=\left({\Omega}_1,\dots ,{\Omega}_j,\dots ,{\Omega}_{N-1}\right)$ as follows:\\

\begin{equation} \label{4}
   L\!\left(\rho\right)
   =
   L\left(
         \sqrt{
            {{\mathbf r}_1}^2+\dots+{{\mathbf r}_j}^2+\dots+{{\mathbf r}_N}^2
         }
   \right)
   =
   O\!\left(N\right)
\;\;\;\;  ,
\end{equation}
\smallskip

\begin{equation} \label{5}
   L\!\left(\Omega_j\right)_{j<(N-2)}
   =
   L\left(
         \arccos{
            \frac{{\mathbf r}_j}{
                                           \sqrt{{{\mathbf r}_1}^2+\dots+{{\mathbf r}_j}^2+\dots+{{\mathbf r}_N}^2}
                                         }
        }
   \right)
   =
   O\!\left(N\right)
\;\;\;\;  .
\end{equation}
\smallskip

\noindent The numbers of elementary operations required to evaluate the wave functions $\phi\!\left(\rho \right)$ and $\Upsilon\!\left({\Omega}\!\right)=\upsilon_1({\Omega}_1) \dots \upsilon_j({\Omega}_j) \dots \upsilon_{N-1}({\Omega}_{N-1})$ at the calculated values of the coordinates $\rho$ and $\Omega$ are finite as $\phi\!\left(\rho \right)$ and $\upsilon_j({\Omega}_j)$ can be expressed in the finite (and not dependent on $N$) numbers of some previously known (elementary) functions of $\rho$ and ${\Omega}_j$. Hence, we find that\\

\begin{equation} \label{6} 
   {\mit cost}\!\left(\!\Psi\!\right)
   =
   {\mit poly}\!\left(N\right)
\;\;\;\;  .
\end{equation}
\smallskip

\noindent This implies that for a system of $N$ particles, in which disjoint pairs interact by arbitrary two-particle potentials, the verification complexity $L$ is upper-bounded by a polynomial\\

\begin{equation} \label{7} 
   L\!\left(H\Psi\!\left(\rho,\Omega\right)\right)
   \le
   {\mit poly}\!\left(N\right)
\;\;\;\;  .
\end{equation}
\smallskip

\noindent The corollary to this conclusion is that for a very wide class of non-relativistic many-body systems, the decision form of the Schrödinger equation is in NP, i.e., in the complexity class of computational problems whose solutions can be verified in polynomial time.\\

\section{Coupled spin systems}

\noindent Let us now turn to coupled spin systems, which can be characterized with the Hamiltonian $H_c=H+H_{int}$ , where the first part $H$ contains all the interactions described by the distances $d_{jk}$ between the $j$ and the $k$ constituent particles of the system, whereas the second part $H_{int}$  contains coupling to an external magnetic field as well as coupling between spins of the constituent particles. We will assume that the system characterized with the Hamiltonian $H_c$ is in the ground state of $H$ at zero energy.\\

\noindent Consider the problem of the zero ground state energy of the Hamiltonian function $H\!\left({\sigma}\!_1,\!\dots\! ,{\sigma}\!_j,\!\dots\! ,{\sigma}\!_{N}\!\right)$ that describe the energy of configuration of a set of $N$ spins $\sigma\!{_j}\hbar=2m\!{_j}\in\left\{\!-\hbar, +\hbar\right\}$  in classical Ising models of a spin glass \cite{Fischer,Guerra}\\

\begin{equation} \label{8} 
   H\!\left({\sigma}_1,\dots ,{\sigma}_j,\dots ,{\sigma}_N\!\right) 
   =
  -\sum_{j<k}{J_{jk}\sigma\!{_j}\sigma\!{_k}}
  -\mu\sum^N_{j}{h_{j}\sigma\!{_j}}
\;\;\;\; ,
\end{equation}
\smallskip

\noindent where real numbers $J_{jk}$ are coupling coefficients, $h\!{_j}$ are external magnetic fields and $\mu$ is the magnetic moment. Namely, does the ground state of the Ising Hamiltonian (\ref{8}) have zero energy?\\

\noindent Since the generic algorithm $A\!\left({\Phi }_{\Psi }\!\right)$ can exactly solve the Schrödinger equation for all Hamiltonians, it can also solve the Schrödinger equation $H_c\!\!\left.\left|\psi \!\right.\right\rangle\!=\!0$ for the Hamiltonian $H_c=H+H\!({\sigma}_1^z,\dots ,{\sigma}_j^z,\dots ,{\sigma}_N^z\!)$, where spins ${\sigma}_j$ in the classical Ising Hamiltonian (\ref{8}) are simply replaced by quantum operators – Pauli spin-1/2 matrices ${\sigma}_j^z$. As it is readily to observe, the decision problem of the Schrödinger equation $H_c\!\!\left.\left|\psi \!\right.\right\rangle\!=\!0$ (i.e., does this equation have a solution?) is in the complexity class NP.\\

\noindent Let $\Psi(\rho,\Omega,m_c)$ denote the exact solution to this equation in the position-spin basis. As it has been just demonstrated, the decision problem of the Schrödinger equation $H\Psi(\rho,\Omega,m_c)=0$ for the non-coupled spin Hamiltonian $H$ is in NP. Regarding the coupling Hamiltonian $H\!({\sigma}_1^z,\dots ,{\sigma}_j^z,\dots ,{\sigma}_N^z\!)$, the validity of a positive answer (i.e., there is a spin configuration $m_c=\left({m}_{c1},\dots ,{m}_{cj},\dots ,{m}_{cN}\right)$ with zero energy) can be tested on a deterministic machine by calculating the Hamiltonian function (\ref{8}) of that configuration $m_c$ in polynomial time. Hence, the complexity of verification $L\left(H_c\!\!\left.\left|\psi \!\right.\right\rangle\!\right)$\\

\begin{equation} \label{9} 
   L\!\left(H_c\!\!\left.\left|\psi \!\right.\right\rangle\!\right)
   =
    L\!\left(
       H\Psi(\rho,\Omega,m_c), H\!({\sigma}_1^z,\dots ,{\sigma}_j^z,\dots ,{\sigma}_N^z\!)\Psi(\rho,\Omega,m_c)
    \right)
\;\;\;\;   
\end{equation}
\smallskip

\noindent is polynomial.\\

\noindent As indicated by the paper \cite{Lucas}, all “the famous” NP problems (such as Karp's 21 NP-complete problems \cite{Karp,Garey}) can be written down as the Ising Hamiltonian (\ref{8}) with only a polynomial number of steps (to be exact, with a polynomial number of spins which scales no faster than $N^3$). Therefore, in just a polynomial number of steps one can get from any NP-complete problem to the problem of the zero ground state energy $H\!\left({\sigma}\!_1,\!\dots\! ,{\sigma}\!_j,\!\dots\! ,{\sigma}\!_{N}\!\right)=0$ of the Ising Hamiltonian (\ref{8}). On the other hand, the generic algorithm $A\!\left({\Phi }_{\Psi }\!\right)$ can find whether the Schrödinger equation $H\!({\sigma}_1^z,\dots ,{\sigma}_j^z,\dots ,{\sigma}_N^z\!)\Psi(\rho,\Omega,m)=0$ with the quantum version of the Ising Hamiltonian (\ref{8}) has a solution and – as a result – resolve the NP-complete problem of interest.\\

\noindent Consequently, we get to the following conclusion: As an arbitrary NP problem is polynomial-time reducible to any NP-complete problem and subsequently to the decision problem of the Schrödinger equation with the quantum version of the Ising Hamiltonian (\ref{8}) that encodes the given NP-complete problem, any problem in the complexity class NP can be exactly solved by the generic algorithm $A\!\left({\Phi }_{\Psi }\!\right)$ with only polynomially more work.\\

\noindent This conclusion means that if the generic algorithm $A\!\left({\Phi }_{\Psi }\!\right)$ were efficient (i.e., polynomial in $N$), then the complexity class NP would be equal to P, the class of computational problems solvable in polynomial time. However, as it is now prevalently believed \cite{Gasarch}, P$\ne$NP, and so it is almost certainly that neither $A\!\left({\Phi }_{\Psi }\!\right)$ nor any other method of the family ${\Phi }_{\Psi }$ can be polynomial in $N$.\\

\section{The exact generic algorithm $A\!\left({\Phi }_{\Psi }\!\right)$ versus brute force}

\noindent Suppose the conjecture P$\ne$NP is true. Then the question naturally arises as to what kind of super-polynomial running time is possible for the generic algorithm $A\!\left({\Phi }_{\Psi }\!\right)$ if we compare it to brute force. On the basis of postulates of quantum mechanics (specifically, the postulate that the Hilbert space ${\mathcal H}$ for the composite system containing two subsystems is the tensor product ${\mathcal H}={\mathcal H}_1\!\otimes{\mathcal H}_2$ of the Hilbert spaces ${\mathcal H}_1$ and ${\mathcal H}_2$ for two constituent subsystems), when solving the Schrödinger equation for a system of $N$ particles the brute-force method will run exponentially in $N$. So, is it possible that $A\!\left({\Phi }_{\Psi }\!\right)$ is \textit{significantly faster} than brute force?\\

\noindent Assume that $A\!\left({\Phi }_{\Psi }\!\right)$ is a sub-exponential time algorithm. Since any NP-complete problem – including the 3-SAT problem – can be written down as the decision problem of the Schrödinger equation for the quantum version of the Ising Hamiltonian (\ref{8}), it follows that the generic algorithm $A\!\left({\Phi }_{\Psi }\!\right)$ can solve any NP-complete problem in sub-exponential time. Yet, as laid down by the widely believed conjecture called the Exponential Time Hypothesis (ETH), the 3-SAT problem does not have a sub-exponential time algorithm \cite{Impagliazzo,Woeginger}. Hence, if the runtime complexity of $A\!\left({\Phi }_{\Psi }\!\right)$ were sub-exponential in $N$, then ETH could be shown to be false. This would imply that many computational problems known to be solved in time $O^*(2^N)$ (such as CHROMATIC NUMBER on an $N$-vertex graph, HITTING SET over an $N$-element universe, TRAVELING SALESMAN problem on $N$ cities and so on) can be improved to $O^*(c^N)$ with some $c<2$ (where the $O^*$ notation is used that suppresses factors polynomial in $N$). However, such an improvement would be highly surprising since the lower bound $O^*(2^N)$ is tight: there is strong evidence that this lower bound matches the running time of the best possible algorithms for those problems \cite{Lokshtanov}.\\

\noindent Thus, most likely, no method of the family ${\Phi }_{\Psi }$ (of generic algorithms capable of exactly solving the Schrödinger equation for an arbitrary physical Hamiltonian $H$) could be significantly faster than the brute-force procedure of merely generating and testing all possible candidate solutions of this equation.\\

\section{Solving the Schrödinger equation for a macroscopic system}

\noindent In order to be able to meaningfully talk about the state of a macroscopic system within the formalism of quantum theory, explicitly, as being described by a certain wave function obeying the Schrödinger equation with a particular Hamiltonian $H\!_M$, it is important to bear in mind the following: Such a wave function can be only an \textit{exact (analytical)} solution of Schrödinger's equation obtained by some \textit{generic} algorithm capable of exactly solving this equation with any physical Hamiltonian. Let us show this.\\

\noindent The exact solution to the Schrödinger equation for a given physical system conveys the most complete information that can be known about the system. In opposition, when solving Schrödinger's equation numerically even small round-off errors in the coefficients of the characteristic polynomial (for the matrix representing the system Hamiltonian $H$) can end up being a large error in the eigenvalues and hence in the eigenvectors. So, in a sense numerical solutions of Schrödinger's equation can be viewed as the loss of the complete information theoretically possible about the system analogous to the loss of the information from the system into the environment in the models of environmental-induced decoherence \cite{Healey,Schlosshauer14}. Thus, if one hopes to make headway in foundational matters, say, to resolve the problem of macroscopic quantum superpositions – linear combinations of the solutions to the macroscopic system's Schrödinger equation – one has to consider only the exact solutions. If not, just by averaging over the possible span errors of the numerical solutions one may get the complete loss of coherence of the phase angles between the numerically acquired elements of the macroscopic superposition.\\ 

\noindent By virtue of the large (and essentially unchecked) number of microscopic constituent particles in any macroscopic system, an ambiguity in the identification of a macroscopic system's Hamiltonian $H\!_M$ is inevitable. This indicates that including or excluding many microscopic degrees of freedom into or from the macroscopic Hamiltonian $H_M$ would have no bearing to its identification. For example, adding or removing a hundred molecules (accounting for numerous microscopic degrees of freedom) to or from the surface of a laptop would have no relevance to the laptop's properties or functioning and so to the laptop's Hamiltonian.\\

\noindent In accordance with the paper \cite{ZhangJ}, a parameterized form of an arbitrary Hamiltonian $H$ determining a quantum (i.e., microscopic) system can be written as\\

\begin{equation} \label{10} 
   H\!\left(\theta\right)
   =
    \sum^{M(N)}_{j}{\!a_{j}\!\left(\theta\right)\hat{X}\!{_j}}
\;\;\;\;  ,
\end{equation}
\smallskip

\noindent where $\theta$ is a vector consisting of parameters governing the quantum (microscopic) evolution, $a_{j}$ are some known real-valued functions of $\theta$, $\hat{X}\!{_j}$ are known Hermitian operators, and $M(N)$ stands for an integer-valued function of the system constituent particle number $N$ (typically $M \ll N^2-1$). However, one would be hard-pressed to write down a similar parameterized form $H\!_M\!\left(\theta\right)$ for the macroscopic Hamiltonian because the precise identification of microscopic parameters $\theta$ (i.e., microscopic degrees of freedom) governing a macroscopic system's evolution would be impossible.\\

\noindent Therefore, to be able (even in principle) to exactly solve the Schrödinger equation for a macroscopic system one must use an algorithm that is not written in terms specialized to particular microscopic degrees of freedom $\theta$, that is, to a precisely identifiable Hamiltonian $H_M\!\left(\theta\right)$. This means to accomplish such a task one would be forced to use only a generic algorithm, namely, one of the family ${\Phi }_{\Psi }$.\\

\noindent But since – assuming ETH – there is no generic algorithm $\in{\Phi }_{\Psi }$ that can be significantly faster than brute force, this would imply that for a macroscopic system (that can be considered as a system containing roughly Avogadro's number ${N}\!_{A} \approx 10^{24}$ of the constituent microscopic particles) the time needed to exactly solve the Schrödinger equation would be proportional, at the lowest estimate, to the amount of $O(2^{{N}\!_{A}})$ elementary operations, which – whatever the amount of time an elementary operation may possibly take – would exceed the current age of the universe by extremely big orders of magnitude.\\

\noindent At this point, one can object that NP-hardness of the problem of exactly solving the Schrödinger equation for a macroscopic system does not mean that any instance of this problem will take exponential in ${N}\!_{A}$ time, as, strictly speaking, NP-hardness is a worst-case notion. So, it might be possible that for many macroscopic systems (or at least for some of them) finding the exact solutions to Schrödinger's equation would take a relatively short time (at any rate, a time polynomial in ${N}\!_{A}$). However, even if we assume that solving exactly Schrödinger's equation takes exponential time only for a handful of specific macroscopic systems, this nevertheless will mean – given inescapable coupling of such “unlucky” systems to the rest of the universe – that eventually (after some fast coupling period) for all the other macroscopic systems of the universe finding their exact wave functions will take exponential time too.\\

\section{Concluding remarks}

\noindent Thus, assuming ETH, there is no real possibility to exactly solve the Schrödinger equation for macroscopic systems, and consequently, there is no sense in describing them within the formalism of quantum theory.\\

\noindent The obvious deduction from this conclusion would be that the “mysterious” division between the microscopic world governed by quantum mechanics and the macroscopic world that obeys classical physics could be explained by NP-hardness of the problem of exactly solving the Schrödinger equation for a macroscopic system. If ETH holds, we cannot take the exact wave function (or the exact state vector) as a universal description of reality since for macroscopic physical systems such a description would be empty, i.e., without realistically reachable predictive content.\\

\noindent After all, the time complexity of the Schrödinger equation, that is, the amount of time taken to exactly solve this equation for a given system as a function of the system's constituent particle number $N$, could be the very reason that justifies the classical-quantum dualism of the Copenhagen interpretation explaining why macroscopic measuring devices cannot be realistically described by the same equation as the microscopic systems under consideration.\\


\begin{thebibliography}{21}

\bibitem{Faye}\label{Faye}
{Faye J.} \textit{Copenhagen Interpretation of Quantum Mechanics}, The Stanford Encyclopedia of Philosophy (Fall 2008 Edition). Edward N. Zalta (ed.) Available:  http://plato.stanford.edu/archives/fall2008/entries/qm-copenhagen/.

\bibitem{Landsman}\label{Landsman}
{Landsman N.} \textit{Between classical and quantum}. In: Handbook of the Philosophy of Science, pp 417-553. Elsevier, 2007.

\bibitem{Schlosshauer13}\label{Schlosshauer13}
{Schlosshauer M., Koer J., Zeilinger A.} A snapshot of foundational attitudes toward quantum mechanics. \textit{Studies in History and Philosophy of Science Part B}, 44(3):222-230, 2013. arXiv:1301.1069.

\bibitem{Styer}\label{Styer}
{Styer D. et al.} Nine formulations of quantum mechanics. \textit{Am. J. Phys.}, 70(3), March 2002.

\bibitem{Vaidman}\label{Vaidman}
{Vaidman L.} Quantum Theory and Determinism. arXiv:1405.4222. 2014.

\bibitem{Penrose}\label{Penrose}
{Penrose R.} On the Gravitization of Quantum Mechanics 1: Quantum State Reduction. \textit{Foundations of Physics}, 445:557-575, 2014.

\bibitem{Paterson}\label{Paterson}
{Paterson M. and Stockmeyer L.} On the number of nonscalar multiplications necessary to evaluate polynomials, \textit{SIAM J. Comput.}, 2(1), 1973.

\bibitem{Baur}\label{Baur}
{Baur W. and Strassen V.} The complexity of partial derivatives. \textit{Theor. Comp. Sc.}, 22:317-330, 1983.

\bibitem{ZhangR}\label{ZhangR}
{Zhang R. and Deng C.} Exact solutions of the Schrödinger equation for some quantum-mechanical many-body systems. \textit{Phys. Rev. A}, 47(1):71-77, 1993.

\bibitem{Fischer}\label{Fischer}
{Fischer K. and Hertz J.} \textit{Spin Glasses.} Cambridge University Press, 1991.

\bibitem{Guerra}\label{Guerra}
{Guerra F. and Toninelli F.} The thermodynamic limit in mean field spin glass models. \textit{Communications in Math. Phys.}, 230 (1):71-79, 2002.

\bibitem{Lucas}\label{Lucas}
{Lucas A.} Ising formulations of many NP problems. \textit{Frontiers in Physics}, 2(5):1-15, 2014.

\bibitem{Karp}\label{Karp}
{Karp R.} \textit{Reducibility Among Combinatorial Problems.} In: Miller R. and Thatcher J. (editors). Complexity of Computer Computations. New York: Plenum. pp. 85–103, 1972.

\bibitem{Garey}\label{Garey}
{Garey M. and Johnson D.} \textit{Computers and Intractability: a Guide to the Theory of NP-Completeness.} New York: Freeman \& Co, 1979.

\bibitem{Gasarch}\label{Gasarch}
{Gasarch W.} P=?NP poll. \textit{SIGACT News}, 33(2):34-47, 2002. Available: http://www.cs.umd.edu/$\sim$gasarch/papers/poll.pdf.

\bibitem{Impagliazzo}\label{Impagliazzo}
{Impagliazzo R., Paturi R., Zane F.} Which problems have strongly exponential complexity? \textit{J. Comput. System Sci.}, 63:512-530, 2001.

\bibitem{Woeginger}\label{Woeginger}
{Woeginger G.} \textit{Exact Algorithms for NP-hard Problems: A Survey.} Combinatorial Optimization - Eureka, You Shrink! Springer-Verlag, pp. 185-207, 2003.

\bibitem{Lokshtanov}\label{Lokshtanov}
{Lokshtanov D., Marx D., Saurabh S.} Lower bounds based on the exponential time hypothesis. \textit{Bulletin of the EATCS}, 84:41-71, 2011.

\bibitem{Healey}\label{Healey}
{Healey R.} Quantum decoherence in a pragmatist view: Resolving the measurement problem. arXiv:1207.7105. 2012.

\bibitem{Schlosshauer14}\label{Schlosshauer14}
{Schlosshauer M.} \textit{The quantum-to-classical transition and decoherence.} In: Aspelmeyer M., Calarco T., Eisert J., Schmidt-Kaler F. (eds.), Handbook of Quantum Information. Springer: Berlin/Heidelberg, 2014.

\bibitem{ZhangJ}\label{ZhangJ}
{Zhang J. and Sarovar M.} Quantum Hamiltonian identification from measurement time traces. arXiv:1401.5780. 2014.\\

\end{thebibliography}
\end{document}